%Paper: 9110059
%From: THEORY%V1.PH.QMW.AC.UK@VTVM2.CC.VT.EDU
%Date: Mon, 21 OCT 91 13:16:54 GMT

\input phyzzx.tex

%this is an input to write the page number at the top of the pages.
%\input topnum.tex

\def\sqr#1#2{{\vcenter{\hrule height.#2pt
      \hbox{\vrule width.#2pt height#1pt \kern#1pt
        \vrule width.#2pt}
      \hrule height.#2pt}}}

\def\gtorder{\mathrel{\raise.3ex\hbox{$>$}\mkern-14mu
             \lower0.6ex\hbox{$\sim$}}}
\def\ltorder{\mathrel{\raise.3ex\hbox{$<$}\mkern-14mu
             \lower0.6ex\hbox{\sim$}}}
\def\dalemb#1#2{{\vbox{\hrule height .#2pt
        \hbox{\vrule width.#2pt height#1pt \kern#1pt
                \vrule width.#2pt}
        \hrule height.#2pt}}}

\def\np{Nucl. Phys.}
\def\pl{Phys. Lett.}
\def\pr{Phys. Rev.}

\def\mpl{Mod. Phys. Lett.}

\REF\dirac{ P. A. M. Dirac, Canad. J. Math. {\bf 2} (1950) 129;
                            \pr\ {\bf 114} (1959) 924 and
                            {\it\lq Lectures on Quantum Mechanics'},
                              Yeshiva University, New York, 1964.}
\REF\teitel{L. Brink, M. Henneaux  and C. Teitelboim, \np\
            {\bf B293} (1987) 505.}
\REF\senjan{P. Senjanovic, Ann. of Phys. {\bf 100} (1976) 227.}
\REF\casal{R. Casalbuoni, \pl\ {\bf B62} (1976) 49, Nuovo Cimento {\bf A33}
            (1976) 389; P. G. O. Freund, unpublished, as quoted in A. Ferber,
             \np\ {\bf B132} (1978) 55.}
\REF\brink{L. Brink and J. H. Schwarz, \pl\ {\bf B100} (1981) 310.}
\REF\greensch{M. B. Green and J. H. Schwarz, \pl\ {\bf B136} (1984) 367.}
\REF\recent{U. Lindstr\" om, M. Ro\v cek, W. Siegel, P. van Nieuwenhuizen and
            A. E. van de Ven, \pl\ {\bf B224} (1989) 285; W. Siegel,
            {\it\lq Proc. of the 1989 Texas A + M String Workshop'};M. B.
            Green and C. M. Hull, \pl\ {\bf B225} (1989) 57; R. E. Kallosh,
            \pl\ {\bf B224} (1989) 273 and \pl\ {\bf B225} (1989) 49;
            S. J. Gates, M. T. Grisaru, U. Lindstr\" om, M. Ro\v cek,
            W. Siegel, P. van Nieuwenhuizen and A. E. van de Ven,
            \pl\ {\bf B225} (1989) 44.}
\REF\greenn{M. B. Green and C. M. Hull,
            {\it\lq Proc. of the 1989 Texas A + M String Workshop'} and
            \np\ {\bf B344} (1990) 115.}
\REF\bengs{I. Bengtsson, \pr\ {\bf D39} (1989) 1158.}
\REF\purely{W. Siegel, \pl\ {\bf B205} (1988) 257; \np\ {\bf B263} (1985) 93;
            and in {\it \lq Unified String Theories'},
            M. B. Green and D. J. Gross eds. (World Scientific, Singapore).}
\REF\batalin{I. A. Batalin and G.A. Vilkovisky, \pl\ {\bf B102} (1981) 27
                        and \pr\ {\bf D28} (1983) 2567.}
\REF\beaulieu{L. Beaulieu, Phys. Rep. {\bf 129} (1985) 1.}
\REF\armonic{E. Nissimov, S. Pacheva and a S. Solomon, \np\
            {\bf B296} (1988) 462; {\bf B297} (1988) 349;
            R. E. Kallosh and M. A. Rahmanov, \pl\ {\bf B209} (1988) 233.}
\REF\howe{E. Sokatchev, \pl\ {\bf B169} (1986) 209 and Class. Quantum Grav.
            {\bf 4} (1987) 237; A. S. Galperin, P. S. Howe and K. S. Stelle,
            \lq{\it The Superparticle and the Lorentz Group}', preprint
            IMPERIAL/90-91/16.}
\REF\huq{M. Huq, \np\ {\bf B315} (1989) 249.}
\REF\sige{W. Siegel, Class. Quantum Gravity {\bf 2} (1985) L95.}
\REF\green{M. B. Green and C. M. Hull, \mpl\ {\bf A18} (1990) 1399.}
\REF\kallosh{R. E. Kallosh, \pl\ {\bf B195} (1987) 369; R. E. Kallosh,
            A. Van Proeyen and W. Troost, \pl\ {\bf B212} (1988) 428.}
\REF\kalloshe{E. A. Bergshoeff and R. Kallosh, \pl\ {\bf B240} (1990) 105;
            E. A. Bergshoeff, R. Kallosh and A. van Proeyen, preprint
            CERN-TH-6020/91.}
\REF\fish{F. Bastianelli, G. W. Delius and E. Laenen, \pl\
            {\bf B229} (1989) 223; J. M. L. Fisch and M. Henneaux,
            {\it\lq A note on the covariant BRST quantization of the
            superparticle'}, Universit\'e Libre de Bruxelles preprint
            ULB-TH2/89-04.}
\REF\miko{A. Mikovi\' c, M. Ro\v cek, W. Siegel, A. E. van de Ven,
           P. van Nieuwenhuizen and J.P. Yamron, \pl\ {\bf B235} (1990) 106.}
\REF\sigee{F. E$\beta$ler, E. Laenen, W. Siegel and J. P. Yamron, \pl\
           {\bf B254} (1991) 411; F. E$\beta$ler, M. Hatsuda, E. Laenen,
           W. Siegel, J. P. Yamron, T. Kimura, A. Mikovi\' c,
           {\it\lq Covariant Quantization of the First-Ilk Superparticle'},
           Stony Brook Preprint ITP-SB-90-77.}
\REF\kalle{R. E. Kallosh, \pl\ {\bf B251} (1990) 134;E. A. Bergshoeff,
           R. Kallosh and A. van Proeyen, \pl\ {\bf B251} (1990) 128;
            M. Huq, \lq{\it Covariant quantization of the broken twisted N=2
           Superparticle}', preprint DOE-ER40200-257.}
\REF\sigehe{U. Lindstr\" om, M. Ro\v cek, W. Siegel, P. van Nieuwenhuizen,
           A. E. van de Ven, J. Math. Phys. {\bf 31} (1990) 1761.}
\REF\anomal{P. S. Howe, U. Lindstr\" om and P. White,
           \pl\ {\bf B246} (1990) 430; W. Troost, P. van Nieuwenhuizen,
           A. van Proeyen, \np\ {\bf B333} (1990) 727.}
\REF\book{W. Siegel, {\it\lq Introduction to String Field Theory'},
           World Scientific Publishing, Singapore, 1988.}

\def\half{{\textstyle {1\over 2}}}

\def\gamu{\gammaˆ\mu}
\def\pmu{pˆ\mu}
\def\psh{\rlap{/}{p}}

\def\intt {\int\!\! }
\def\Fa{\PhiˆA}
\def\aFa{\Phi_Aˆ\star}
\def\aFb{\Phi_Bˆ\star}
\def\dlf {\partial_{l}}
\def\dri {\partial_{r}}
\def\tilthet{\tilde\theta}
\def\stilthet{{\tilde\theta}ˆ\star}
\def\stheta{\thetaˆ\star}
\def\sxmu{x_\muˆ\star}
\def\sg{gˆ\star}
\def\sd{dˆ\star}
\def\slambda{\lambdaˆ\star}
\def\spsi{\psiˆ\star}
\def\scc{cˆ\star}
\def\tilpi{\tilde\pi}

\Pubnum = {QMW-91-18}
\date = {June 1991}
\titlepage
\title {\bf ON THE COVARIANT QUANTIZATION OF THE 2nd. ILK SUPERPARTICLE.}

\author {J. L. V\'azquez-Bello\footnote\S{Work supported by
                                                   CONACYT-MEXICO.}}

\address {Physics Department, Queen Mary and Westfield College,\break
             Mile End Road, London E1 4NS,\break
                 UNITED KINGDOM.}

%\endpage
%\nopagenumbers

\abstract {This paper is devoted to the quantization of the second-ilk
superparticle using the Batalin-Vilkovisky method. We show the full structure
of the master action. By imposing gauge conditions on the gauge fields rather
than on coordinates we find a gauge-fixed quantum action which is free.
The structure of the BRST charge is exhibited, and the BRST cohomology yields
the same physical spectrum as the light-cone quantization of the usual
superparticle.}

\endpage
%\baselineskip = 15pt
%\pagenumber=1
%\sequentialequations

\chapter{Introduction.}

Constraints of a dynamical system are classified as first and second class,
according to their Poisson bracket relations [\dirac ,\senjan]. However, the
mixing of first and second class constraints, and the difficulty of their
separation in a covariant way has proved to be a problem in the covariant
quantization of superparticles and strings [\teitel ,\bengs ].
%Therefore, models which possess mixed constraints should be studied more
%carefully before performing their quantization [\senjan ].

Covariant quantization of the original superparticle [\casal ,\brink ],
together with its superstring generalizations, has proved problematic because
of the mixing of first and second class constraints [\recent -\purely ].
Progress has been made in the covariant quantization of the
superparticle using the Batalin and Vilkovisky methods (BV) which can be
applied in the presence of second-class constraints [\batalin ,\beaulieu ], or
using harmonic variables for the separation of the fermionic constraints into
first and second class [\armonic ,\howe ]. However, the latter approach
suffers from some non-locality problems. Harmonic variables are additional
bosonic variables added to the usual bosonic and fermionic coordinates.
Appropriate first class constraints are imposed on the harmonic variables to
assure that they are purely gauge degrees of freedom [\huq ].
Following this idea of adding variables to separate constraints, an alternate
possibility is the introduction of fermionic coordinates to render all
fermionic constraints first class. Siegel has shown that by introducing a
momentum conjugate to the fermionic variable $\theta$ and a gauge field for
the fermionic world-sheet symmetry, one can obtain systems with purely first
class constraints [\purely ,\sige ]. However, although several alternatives
have been proposed to address the quantization problem of the superparticle,
which are based on modifications to the formulation given by Siegel [\sige ],
none of them has led so far to a satisfactory solution, essentially because a
suitable gauge-fixing fermion has not been found or the BRST operator does
not give the correct cohomology [\recent ,\kallosh -\fish ,\sigehe ].
To avoid some of the issues involved in the covariant quantization of those
formulations, further modifications have been proposed. In [\miko ,\sigee ],
two first-class formulations have been proposed, the first-ilk and the
second-ilk superparticle. A similar model was proposed in [\kalle ]. Both of
them allow covariant quantization. Another proposal was made in [\green ]
where the modified action involves additional fermionic coordinates. In each
case, the BRST cohomology gives the spectrum of N=1 super Yang-Mills.

The treatment of the quantum second-ilk superparticle in [\miko ] was
incomplete and the purpose of this paper is to complete the analysis, using
the methods of BV. We explicitly show all the relevant steps in the
calculation of the master action and the BRST charge, correcting the form of
the master action given in [\miko ]. There, a BRST charge for the second-ilk
superparticle was written down and one of our aims here is to compare this
with the BRST charge that arises in the BV approach.
%where contributions to the anti-commuting spinors
%$\theta_0 ,\dots ,\theta_{2n} ,\dots$ were not included in the BRST
%transformation for $xˆ\mu$.
Section 2 reviews the original superparticle action, the second-ilk
superparticle action and their symmetries. In section 3, we analyse the ghost
structure which provides a representation of the BRST algebra to find the
minimal set of fields which enter in the BV procedure. In section 4, the
master action of the BV method is obtained and is used in section 5 to
determine the quantum action and the BRST charge.

\chapter {Classical Actions and Symmetries.}

The original superparticle (which we shall refer to as SSP0) described by
an action with mixed first and second class constraints was formulated in
[\casal ,\brink ],
%first studied by Casalbuoni, Freund [\casal ], and Brink-Schwarz [\brink ],
and generalized to superstrings by Green and Schwarz [\greensch ].
The evolution of the SSP0 superparticle is represented by a world-line in
ten-dimensional superspace ($xˆ\mu (\tau ),\theta (\tau )$) parameterized by
$\tau$, where $\mu =0,...,9$ and $\theta$ is an anti-commuting Majorana-Weyl
spinor. The SSP0 action is given by
$${\sl S}_{SSP0} =\intt d\tau\bigl[p_\mu ({\dot x}ˆ\mu
                               -i\bar\theta\gamu\dot\theta )
                                       -\half epˆ2\bigr]\eqn\original$$
where $\dot\theta=d\theta /d\tau$, $pˆ\mu$ is the momentum and $e$ is the
einbein on the world-line. The SSP0 action is invariant under rigid
space-time supersymmetry transformations together with world-line
reparameterizations and a local fermionic symmetry, although there is no gauge
field for this symmetry. In a covariant quantization it is necessary to find a
covariant gauge choice for the fermionic symmetry. As there is no gauge field
for the local fermionic symmetry, this can only be fixed by imposing
conditions on $(xˆ\mu ,\theta ,e )$. There have been numerous attempts to find
a covariant quantization of the SSP0 superparticle given by \original , but
there is no satisfactory covariant gauge choice [\greenn ].
%which describe the massless sector of the superstring [\bengs ,\huq ].
% However,
 a
%covariant quantization of the SSP0 action using the methods of BV yields
%unsatisfactory results [\fish ].

In this paper, we present the results of the covariant quantization of a
further modification of the superparticle, the second-ilk superparticle
of [\miko ]. The second-ilk superparticle has only first-class constraints.
This new superparticle action is formulated in a superspace
with coordinates $(xˆ\mu ,\theta_0 ,\dots ,\theta_{2n} ,\dots )$, where
$\theta_0 ,\dots ,\theta_{2n},\dots$ are anti-commuting spinors. The action is
$$\eqalign{S_0 &=p_\mu{\dot x}ˆ\mu -gpˆ2 -i\psi_1\psh (d_0 -2\psh\theta_0 )
        +i\sum_{n=0}ˆ{+\infty}{\dot\theta}_{2n} (d_{2n}-\psh\theta_{2n})\cr
        &\qquad -\sum_{n=0}ˆ{+\infty}\lambda_{2n+1} (d_{2n} +d_{2n+2}
                                                      -2\psh\theta_{2n+2}).
\cr}\eqn\class$$
and
is invariant under global space-time supersymmetry and a number of
local symmetries. These symmetries are given by
$$\eqalign{&\delta\theta_0 =\kappa\psh -i\theta_1 ,\cr
           &\delta xˆ\mu   =2\xi\pmu +i\kappa\bigl[\gamu (d_0 -2\psh\theta_0 )
                                     -\psh\gamu\theta_0\bigr] \cr
                   &\qquad +\sum_{n=0}ˆ{+\infty}\theta_{2n+1}\gamu
                                     (\theta_{2n} -\theta_{2n+2}) ,\cr
                &\delta g  =\dot\xi +2i\psi_1\psh\kappa ,\cr
     &\delta\lambda_{2n+1} ={\dot\theta}_{2n+1} ,\cr
             &\delta\psi_1 =\dot\kappa ,\cr
        &\delta\theta_{2n} =-i(\theta_{2n+1} +\theta_{2n-1}) ,\cr
            &\delta d_{2n} =-2i\theta_{2n+1}\psh .
\cr}\eqn\symmet$$
The
gauge fields $g$, $\psi_1$ and $\lambda_{2n+1}$ are all lagrange multipliers
imposing the infinite set of constraints
$$pˆ2 =0, \qquad \psh d_0 =0,
         \qquad d_{2n} +d_{2n+2} -2\psh\theta_{2n+2} =0.\eqn\constr$$
The momentum
$pˆ\mu$ is an auxiliary field whose algebraic equation of motion is given by
$$\pmu ={1\over {2g}}(4i\psi_1\pmu\theta_0 +{\dot x}ˆ\mu -i\psi_1\gamu d_0
        -i\sum_{n=0}ˆ{+\infty}{\dot\theta}_{2n}\gamu\theta_{2n}
        +2\sum_{n=0}ˆ{+\infty}\lambda_{2n+1}\gamu\theta_{2n+2} ).\eqn\moment$$
The
remaining classical field equations are
$${\dot p}ˆ\mu =0, \qquad \psh{\dot\theta}_0 -pˆ2\psi_1 =0,
         \qquad {\dot\theta}_{2n}\psh -i\lambda_{2n-1}\psh =0.\eqn\equats$$

\chapter{The Ghost Structure of the Superparticle.}

For theories in which the classical gauge algebra closes off-shell, it is
straightforward to construct a BRST invariant action. For theories in which
the gauge algebra only closes on-shell, however, the standard BRST approach
do not work and it is convenient to use the BV method to quantize them
[\batalin ,\beaulieu ]. Furthermore, the quantum action constructed using BV
or alternate procedures should be invariant under BRST transformations which
reflect the gauge invariance at the classical level.
The first step towards the covariant quantization of models with open
gauge algebras, is to study the ghost structure in order to find the minimal
set of fields that enter in the BV quantization procedure.
The ghost structure is found by demanding that the minimal set of fields
provide a representation of the BRST algebra.

Now consider the application of the BV method to determine the minimal set
of fields of the second-ilk superparticle model given by \class .
The BRST transformation of any of the classical fields appearing in \class ,
is given by replacing the parameter of the gauge transformation with the
corresponding ghost.  For the action \class , we introduce ghosts
$(c,\tilthet_1 ,\dots ,\tilthet_{n},\theta_{2n+1},\psi_{n})$ corresponding
to the classical symmetries \symmet\ with gauge parameters of opposite
Grassmann parity to the ghost set,
$(\xi ,\kappa ,\dots ,\tilthet_{n},\theta_{2n+1},\psi_{n})$.
Then,
considering the on-shell nilpotency condition of the BRST
transformations on all the classical and ghost fields, we will construct the
ghost spectrum of the superparticle \class\ whose structure can be
represented by infinite towers of ghost fields.

For the $10$-dimensional superparticle defined by the action \class , the BRST
transformations for the classical fields are
$$\eqalign{& s xˆ\mu =2c\pmu
                          +i\tilthet_1\bigl[\gamu (d_0 -2\psh\theta_0 )
                                     -\psh\gamu\theta_0\bigr]  \cr
                     &\qquad +\sum_{n=0}ˆ{+\infty}\theta_{2n+1}\gamu
                                     (\theta_{2n} -\theta_{2n+2}) ,\cr
               & s g  =\dot c+2i\psi_1\psh\tilthet_1
                             +i\psi_2 (d_0 -2\psh\theta_0 ) ,\cr
          & s\theta_0 =\tilthet_1\psh -i\theta_1 ,\cr
    & s\lambda_{2n+1} ={\dot\theta}_{2n+1} ,\cr
           &  s\psi_1 = -({\dot{\tilde\theta}}_1 -\psh\psi_2 ) ,\cr
      &  s\theta_{2n} =-i(\theta_{2n+1} +\theta_{2n-1}) ,\cr
          &  s d_{2n} =-2i\theta_{2n+1}\psh .
\cr}\eqn\owleyes$$
The
BRST transformations for the ghost fields are given by demanding the
nilpotency of the generator $ s $, up to terms that vanish when the classical
equations of motion are satisfied, so that the BRST transformations becomes
nilpotent on-shell. Then
$$\eqalign {& s\tilthet_{n} =(-)ˆ{n+1}\psh\tilthet_{n+1} ,\cr
            &  s\psi_{n+1} = (-)ˆ{n+1} ({\dot{\tilde\theta}}_{n+1}
                            -\psh\psi_{n+2} ) ,\cr
            &  sc =i\tilthet_1\psh\tilthet_1
                   -i\tilthet_2 (d_0 -2\psh\theta_0 ),
\cr}\eqn\xowleyes$$
where $\theta_{2n+1}$ and $\tilthet_1$ are ghosts, while
$\tilthet_2 ,\dots ,\tilthet_{2n+1}$ are ghosts-for-ghosts with Grassmann
parities and space-time chiralities that alternate with the level number.
However, the construction
of the quantum action requires the introduction of some new fields.
For each {\it n}'th generation ghost field, one introduces an
anti-ghost and Nakanishi-Lautrup (NL) fields, plus \lq extra-ghosts'
together with the corresponding extra-NL fields, so that
at the {\it n}'th generation the ghost is supplemented by BRST doublets.
This set of fields is always sufficient to construct a quantum action.
The minimal set of fields that enter in the BV quantization procedure is
determined by the classical gauge symmetries, together with the requirement
that the BRST transformations of the classical fields and the ghosts should be
nilpotent on-shell.  This procedure also fixes much of the structure of the
master action.  Then, the minimal set of fields for the first-class
superparticle \class , based on the classical symmetries \symmet , consists
of the classical and ghost fields introduced above,
$$\Fa_{min} =\{xˆ\mu ,\pmu ,g,c,\theta_0 ,d_0 ,\psi_1 ,\theta_{2n},
             d_{2n},\lambda_{2n+1},\tilthet_{n},\theta_{2n+1},
             \psi_{n+1} \} .\eqn\minfield$$

A common feature of superparticle and superstring models is the
infinite-reducibility of these systems. The existence of an
infinite number of ghost coordinates may seem to be a complication
but they package together into an infinite-dimensional metaplectic
representation of an orthosymplectic supergroup [\greenn ,\book ].

\chapter{BV Quantization}

In this section, the quantization of \class\ will be discussed. We begin by
briefly reviewing the BV procedure for constructing BRST transformations and
the corresponding quantum action, which works for arbitrary systems with open
algebras [\batalin ]. The \lq minimal' set of fields $\Fa$ that enter in the
BV method is determined by the classical gauge symmetries, together with the
requirement that the BRST transformations of the classical fields and the
ghosts should be on-shell nilpotent {\it i.e.}, $sˆ2 =0$ on any field, up to
terms proportional to the equations of motion.
For each field $\Fa$ a corresponding \lq anti-field', $\aFa$, of the opposite
Grassmann parity is introduced. Then, the first step in determining the
quantum action is to find the solution $S(\Fa ,\aFa )$ to the master
equation,
$${{\dri S}\over {\partial\Fa}}{{\dlf S}\over {\partial\aFa}}=0,\eqn\master$$
subject
to the boundary condition that the master action reduces to the
classical action when the \lq anti-fields' are set to zero,
$S(\Fa ,\aFa )\vert_{\aFa =0} = S_0 (\Fa )$. The symbols $r$ and $l$ in
\master\ refer to right and left derivatives respectively, the order
being crucial due to the Grassmann nature of some of the fields.
Then, for any gauge fermion $\Psi (\Fa )$, which is typically a sum of terms
of the form $({\it anti-field})\times ({\it gauge-condition})$, the
corresponding quantum action is found by making the following substitution
for the anti-fields in $S$,
$$\aFa = {{\dlf \Psi}\over {\partial\Fa}},\eqn\antifi$$
to give,
$$S_Q (\Fa )= S(\Fa ,\aFa )
             \bigg\vert_{\aFa = {{\dlf \Psi}\over {\partial\Fa}}}.\eqn\wendy$$
This quantum
action is then invariant under the modified BRST transformations given by
$$\hat s \Fa ={{\dlf S}\over {\partial\aFa}}
               \bigg\vert_{\aFa = {{\dlf \Psi}\over {\partial\Fa}}}\eqn\liza$$
which
are nilpotent up to terms which vanish when the equations of motion
derived from the quantum action $S_Q$ are satisfied. The gauge fermion must be
chosen so as to remove the gauge degeneracy of the classical action and
give invertible kinetic terms. Using \liza\ one can define the generating
functional $W[J]$ as usual via the path integral
$$\exp(iW[J])=\intt [d\Phi ]\exp\{ i(S_Q [\Phi ]+J_i \Phiˆi )\} ,\eqn\linda$$
where $W[J]$ must
be regularised and normalised. The functional integral will then be BRST
invariant using the quantum action \wendy , provided that the measure is BRST
invariant. If not, then one seeks local counterterms to cancel the variation
of the measure, so that the functional integral is BRST invariant. Within the
BV formalism, this corresponds to seeking corrections to the master action of
the form $W = S + {\it 0}(\hbar )$, such that $W$ satisfies
$$\half (W,W) =-i\hbar\Delta W +\hbar a_\nu cˆ\nu
                               + {\cal O}(\hbar ˆ2 )\eqn\violet$$
where $a_\nu$ are the anomalies and $cˆ\nu$ are the ghost fields.
%and $W(\Phi ,\Phiˆ\star )$ is the full master solution to \violet , which
%to zero{\it th} order in $\hbar$ corresponds to $S(\Phi ,\Phiˆ\star )$.
A remarkable result is that if there is no local solution to the modified
master equation \violet , then the theory is anomalous and the quantum theory
is inconsistent. A discussion of anomalies in the BV formalism with an
explicit regularization of the path integral is given in [\anomal ].

Using the BV formalism, we find the solution $S_{min}$ to the master
equation \master\ for the minimal set of fields \minfield , when
expanded in powers of anti-fields, takes the form
$$\eqalign{ S_{min} &= S_0 +S_1 +S_2 \cr
                    &= S_0 +\intt d\tau\aFa (s\Fa )
                         +\half\intt d\tau\aFa\aFb Eˆ{AB}(\Phi ,\Phiˆ\star )
,\cr}\eqn\rubia$$
where
$S_0$ is the classical action, \class .  The term linear in anti-fields
is given by
$$\eqalign{S_1 =\intt &d\tau\Bigl\{
                 \stheta_0 (\psh\tilthet_1 -i\theta_1 )
          -i\sum_{n=1}ˆ{+\infty}\stheta_{2n}(\theta_{2n+1}+\theta_{2n-1}) \cr
      &\quad  -\sum_{n=0}ˆ{+\infty}(-)ˆ{n}\stilthet_{n}\psh\tilthet_{n+1}
                    +\sxmu\sum_{n=0}ˆ{+\infty}\theta_{2n+1}\gamu
                               (\theta_{2n} -\theta_{2n+2}) \cr
             &\quad +\sxmu\Bigl( 2c\pmu
                        +i\tilthet_1\bigl[\gamu (d_0 -2\psh\theta_0 )
                              -\psh\gamu\theta_0\bigr]\Bigr) \cr
     &\quad -2i\sum_{n=0}ˆ{+\infty}\sd_{2n}\psh\theta_{2n+1}
                + \sg [\dot c +2i\psi_1\psh\tilthet_1
                               +i\psi_2 (d_0 -2\psh\theta_0 )] \cr
     &\quad +\sum_{n=0}ˆ{+\infty}\slambda_{2n+1}{\dot\theta}_{2n+1}
                + \sum_{n=1}ˆ{+\infty}(-)ˆ{n}\spsi_{n}
                         ({\dot{\tilde\theta}}_{n}-\psh\psi_{n+1}) \cr
    &\quad +i\scc [\tilthet_1\psh\tilthet_1 -\tilthet_2 (d_0 -2\psh\theta_0 )]
\Bigr\} ,\cr}\eqn\patsy$$
while
the term quadratic in anti-fields is
$$\eqalign{S_2 =\intt d\tau &\Bigl[\sg\stheta_0\tilthet_2
                      -\sg\sum_{n=0}ˆ{+\infty}\stilthet_{n}\tilthet_{n+2}
                             -\sg\sum_{n=0}ˆ{+\infty}\spsi_{n}\psi_{n+2} \cr
      &\quad +4i\sg\scc\tilthet_1\tilthet_2
                    -i\sg\sxmu (\tilthet_1\gamu\tilthet_1
                                   +\tilthet_2\gamu\theta_0 )  \cr
      &\quad -\sxmu\sum_{n=1}ˆ{+\infty}\spsi_{n}\gamu\tilthet_{n+1}
                    -\scc\sum_{n=1}ˆ{+\infty}(-)ˆ{n}\spsi_{n}\tilthet_{n+2}
\Bigr].\cr}\eqn\sexy$$
This
minimal action \rubia\ corrects the one given in [\miko ]. The $\sxmu$ and
$\scc$ terms in $S_1$ and the term $\sg\scc$ term in $S_2$ differ from
these of [\miko ].
%In [\miko ],
%the corresponding BRST transformation for $xˆ\mu$ is incomplete. There, they
%did not include contributions to the anti-commuting spinors
%$\theta_0 ,\dots\theta_{2n} ,\dots$, so that the structure of the master
%action is modified in the antifields $\sxmu$ and $\scc$ in $S_1$ and in the
%antifields $\sg$, $\scc$ in $S_2$.
The full master action is then given by
adding to $S_{min}$ the non-minimal terms $S_{non-min}$, where antighost
fields ${\hat c}ˆ\star$, ${\hat{\tilthet}}_{n}ˆ\star$,
${\hat\theta}_{2n+1}ˆ\star$ together with the corresponding
NL fields $\pi$, $\tilpi_{n}$, $\pi_{2n+1}$ are required, so that
at the {\it n}'th generation the ghost fields are supplemented by {\it n}
BRST doublets. The non-minimal term is then
$$S_{non-min} ={\hat c}ˆ\star\pi
        +\sum_{n=1}ˆ{+\infty}{\hat{\tilthet}}_{n}ˆ\star\tilpi_{n}
        +\sum_{n=0}ˆ{+\infty}{\hat\theta}_{2n+1}ˆ\star\pi_{2n+1}.\eqn\nonmin$$
There
are terms in the master action which are quadratic in ghost anti-fields,
which cannot be found by solving the master equation \master\ to first order
in anti-fields, so that it is necessary to consider higher order terms in the
anti-fields to find the master action.

\chapter {Quantum Action and BRST-Charge.}

To
define the quantum theory, it is necessary to \lq halve' the extended
configuration space $(\Fa ,\aFa )$ by specifying a hypersurface which
is defined by the condition \antifi , and the corresponding quantum
action $S_Q$ is given by evaluating the master action $S(\Fa ,\aFa )$ on
\antifi\ to give \wendy . However, the gauge fermion in \antifi\ must be
chosen so as to remove the gauge degeneracy of the classical action and
give invertible kinetic terms, so that propagators are well defined.
The gauge fermion typically includes a sum of terms consisting of anti-ghosts
or extra-ghosts times a gauge condition for each of the gauge fields.

We now turn to discuss the choice of gauge. First, the classical gauge
symmetries \symmet\ must be fixed. We shall impose gauge conditions
on the gauge fields $g$, $\psi_{n}$ and $\lambda_{2n+1}$ rather than on
coordinates. The simplest gauge choice is $g=1$, $\psi_{n} =0$ and
$\lambda_{2n+1} =0$, which is implemented by the gauge fermion [\miko ],
$$\Psi (\Fa )=\intt d\tau\Bigl[ (g-1)\hat c
          +\sum_{n=1}ˆ{+\infty}\psi_{n}{\hat{\tilde\theta}}_{n}
            +\sum_{n=0}ˆ{+\infty}\lambda_{2n+1}{\hat\theta}_{2n+1}
          \Bigr],\eqn\gauge$$
where $\hat c$, ${\hat{\tilde\theta}}_{n}$ and ${\hat\theta}_{2n+1}$ are
antighost fields. However, these gauges can only be imposed locally, since
each gauge field should be set equal to a constant {\it modulus} and these
moduli should be integrated over.
It will be convenient to consider a slightly more general class of gauges in
which the gauge fields $g$, $\psi_{n}$ and $\lambda_{2n+1}$ are set equal to
some fixed background fields $\tilde g$, ${\tilde\psi}_{n}$ and
${\tilde\lambda}_{2n+1}$, so that $g=\tilde g$, $\psi_{n}={\tilde\psi}_{n}$
and $\lambda_{2n+1}={\tilde\lambda}_{2n+1}$.
%On choosing
%constant backgrounds $\tilde g =a$, ${\tilde\psi}_{n} =\mu_{n}$
%and $\tilde\lambda_{2n+1} =\beta_{2n+1}$ we recover the previous gauge
%choice. The general gauge will be discussed elsewhere.

Following the standard steps of the BV procedure, we derive a gauge-fixed
quantum action which takes the free form
$$\eqalign {S_Q &=p_\mu{\dot x}ˆ\mu -pˆ2 +(g-1)\pi + \hat c \dot c
                  +i\sum_{n=0}ˆ{+\infty}{\dot\theta}_{2n} d_{2n} \cr
           &\qquad +\sum_{n=1}ˆ{+\infty}(-)ˆ{n}{\hat{\tilde\theta}}_{n}
                           {\dot{\tilde\theta}}_{n}
                   +\sum_{n=0}ˆ{+\infty}{\hat\theta}_{2n+1}
                           {\dot\theta}_{2n+1} \cr
           &\qquad +\sum_{n=1}ˆ{+\infty}\psi_{n}{\tilde\pi}_n
                        +\sum_{n=0}ˆ{+\infty}\lambda_{2n+1}\pi_{2n+1}
,\cr}\eqn\guiness$$
after the following field redefinitions
$$\eqalign{ & d_{2n}' =d_{2n} -\psh\theta_{2n} ,\cr
            &    \pi '=\pi -pˆ2 ,\cr
         & \tilpi_{1}'=\tilpi_1 -i\psh (d_0 -2\psh\theta_0
                                   +2\hat c \tilthet_1 ) ,\cr
       &  \tilpi_{2}' =\tilpi_2 +\psh{\hat{\tilde\theta}}_1
                                   +i\hat c (d_0 -2\psh\theta_0 ) ,\cr
       &  \tilpi_{n}' =\tilpi_{n} +\psh{\hat{\tilde\theta}}_{n-1}
                                   -(-)ˆ{n}\hat c {\hat{\tilde\theta}}_{n-2}
                                   ,\qquad \forall n\geq 3 ,\cr
       &  \pi_{2n+1}' =\pi_{2n+1} -d_{2n} -d_{2n+2}
                                   +2\psh\theta_{2n+2} .\cr}\eqn\lager$$
We have dropped the primes for brevity. Further, the quantum
action \guiness\ can be shown to be invariant under the modified BRST
transformations \liza , which satisfy $\hat s ˆ2 =0$ when the field equations
of motion derived from \guiness\ are satisfied.  As the quantum action defines
a free theory, it is strightforward to quantize it by imposing canonical
commutation relations on the operators corresponding to the variables in
\guiness . The modified BRST transformations for the classical fields are
then
$$\eqalign{&\hat s xˆ\mu =2c\pmu
                          +i\tilthet_1\bigl[\gamu (d_0 -\psh\theta_0 )
                                     -\psh\gamu\theta_0\bigr]
                          +i\hat c (\tilthet_1\gamu\tilthet_1
                                +\tilthet_2\gamu\theta_0 ) \cr
                       &\qquad +\sum_{n=0}ˆ{+\infty}\theta_{2n+1}\gamu
                                     (\theta_{2n} -\theta_{2n+2})
                                -\sum_{n=1}ˆ{+\infty}{\hat{\tilde\theta}}_{n}
                                       \gamu\tilthet_{n+1} ,\cr
            & \hat s g  =\dot c+2i\psi_1\psh\tilthet_1
                             +i\psi_2 (d_0 -\psh\theta_0 )
                             -\sum_{n=1}ˆ{+\infty}{\hat{\tilde\theta}}_{n}
                                                 \psi_{n+2}  ,\cr
  &\hat s\lambda_{2n+1} ={\dot\theta}_{2n+1} ,\cr
      &  \hat s\theta_0 =\psh\tilthet_1 -i\theta_1 +\hat c\tilthet_2 ,\cr
    & \hat s\theta_{2n} =-i(\theta_{2n+1} +\theta_{2n-1}) ,\cr
        & \hat s d_0 =-i\psh\theta_1 -pˆ2\tilthet_1 -\hat c\psh\tilthet_2 ,\cr
       & \hat  s d_{2n} =-i\psh (\theta_{2n+1} -\theta_{2n+1} ) ,\cr
        &  \hat s\psi_1 = -({\dot{\tilde\theta}}_1 -\psh\psi_2
                                              +\hat c\psi_3 ),
\cr}\eqn\newowlsa$$
the
modified BRST transformations for ghost and ghosts-for-ghosts fields are
$$\eqalign{&\hat s\tilthet_{n} =(-)ˆ{n+1}\psh\tilthet_{n+1}
                          +(-)ˆ{n}\hat c \tilthet_{n+2} ,\cr
     & \hat s\psi_{n+1} = (-)ˆ{n+1} ({\dot{\tilde\theta}}_{n+1}
                            -\psh\psi_{n+2} )+(-)ˆ{n+1}\hat c \psi_{n+3} ,\cr
             & \hat s c =i\tilthet_1\psh\tilthet_1
                          -i\tilthet_2 (d_0 -\psh\theta_0 )
                           +4i\hat c \tilthet_1\tilthet_2
                           -\sum_{n=1}ˆ{+\infty}(-)ˆ{n}
                                   {\hat {\tilde\theta}}_{n}\tilthet_{n+2},
\cr}\eqn\newowlsb$$
and the modified BRST transformations for anti-ghosts and NL fields, plus
\lq extra-ghosts' together with the corresponding extra-NL fields are
$$\eqalign{\hat s\hat c &=\pi +pˆ2 ,\cr
\hat s{\hat{\tilde\theta}}_1 &=\tilpi_1 +i\psh (d_0 -\psh\theta_0
                         +2\hat c\tilthet_1 ) ,\cr
\hat s{\hat{\tilde\theta}}_2 &=\tilpi_2 -\psh{\hat{\tilde\theta}}_1
                         -i\hat c (d_0 -\psh\theta_0 ) ,\cr
\hat s{\hat{\tilde\theta}}_{n} &=\tilpi_{n} -\psh{\hat{\tilde\theta}}_{n-1}
                         +(-)ˆ{n} \hat c{\hat{\tilde\theta}}_{n-2} ,\cr
\hat s{\hat\theta}_{2n+1} &=\pi_{2n+1} +d_{2n} +\psh\theta_{2n} +d_{2n+2}
                              -\psh\theta_{2n+2} ,\cr
            \hat s \pi &=0 , \qquad \hat s \pi_{2n+1} =0  ,\cr
     \hat s \tilpi_1 &= -2i\pi\psh\tilthet_1 ,\cr
     \hat s \tilpi_2 &= \psh\tilpi_1 -i(d_0 -\psh\theta_0 )\pi ,\cr
   \hat s \tilpi_{n} &= \psh\tilpi_{n-1} +(-)ˆ{n}\hat c\tilpi_{n-2}
                        +{\hat{\tilde\theta}}_{n-2}\pi .
\cr}\eqn\newowlsc$$
Taking into account the change of variables \lager , the action \guiness\ is
invariant under the BRST transformations generated by the BRST charge
$$\eqalign{Q_{BRST} &=cpˆ2
            -\sum_{n=0}ˆ{+\infty}\theta_{2n+1} (d_{2n} +\psh\theta_{2n} )
            -\sum_{n=0}ˆ{+\infty}\theta_{2n+1}(d_{2n+2}-\psh\theta_{2n+2}) \cr
         &\qquad +i\tilthet_1\psh (d_0 -\psh\theta_0 )
                 -i\hat c\tilthet_2 (d_0 -\psh\theta_0 )
                 +i\hat c\tilthet_1\psh\tilthet_1  \cr
         &\qquad -\sum_{n=1}ˆ{+\infty}{\hat{\tilde\theta}}_{n}
                                                \psh\tilthet_{n+1}
                 -\hat c\sum_{n=1}ˆ{+\infty}(-)ˆ{n}{\hat{\tilde\theta}}_{n}
                                   \tilthet_{n+2} .\cr}\eqn\amor$$
After the following change of
variables
$$d_{2n} \rightarrow t_{2n} ,\qquad
          {\hat{\tilde\theta}}_{n} \rightarrow{\tilde t}_{n},\eqn\tees$$
and using
the following definitions
$$d_{n} =-t_{n} +\psh\theta_{n} ,\qquad
                 q_{n} =-t_{n} -\psh\theta_{n} ,\eqn\variables$$
the BRST charge takes a simple form
$$Q_{BRST} =Q_0 + cpˆ2 - \hat c f ,\eqn\simpleq$$
where
$$Q_0 =-\sum_{n=0}ˆ{+\infty}[{\tilde t}_{n}\psh\tilthet_{n+1}
              -\theta_{2n+1} (q_{2n} +d_{2n+2} )] \eqn\qnut$$
and
$$f =\sum_{n=0}ˆ{+\infty}(-)ˆ{n}{\tilde t}_{n}\tilthet_{n+2}
              -i\tilthet_1\psh\tilthet_1 \eqn\fbit$$
and we have defined ${\tilde t}_0 =id_0$.
Further, the BRST charge \simpleq\ is both conserved and nilpotent, so that
the physical spectrum of the first-class superparticle corresponds to the
cohomology classes of the BRST charge $Q_{BRST}$. Our BRST operator \simpleq\
has exactly the same structure as that of [\miko ], which was computed using
different methods. The BRST cohomology is derived in [\miko ] and yields
the same physical spectrum as the light-cone quantization of the usual
superparticle.

To summarise,  we have used the methods of Batalin and Vilkovisky to
quantize a second-ilk superparticle, \class , which is free of second
class constraints. The BV quantization of this model was also
considered in [\miko ] using a different approach. \footnote*{Although our
expressions for the BRST operator agree, \rubia -\sexy\ correct errors in the
expression for the master action given in [\miko ].}
By solving the BV master equation \master\ and using the gauge
fermion \gauge\ we found a quantum action which, after some field
redefinitions, led to the free quantum action
\guiness , which is invariant under the BRST transformations generated by
the BRST charge \simpleq . It is straightforward to perform an
operator quantization of \class , as in [\miko ], and study the BRST
cohomology to find the physical spectrum, as the quantum action defines a
free theory.
This gives the physical spectrum of the ten-dimensional super Yang-Mills
theory. It should be of interest to study the structure of
second-ilk superparticles and possible generalizations to superstrings, since
at the classical level the evolution of the superparticle is represented by an
infinite set of classical fields which at first generation level in the BRST
transformations involve an infinite set of ghosts which at higher level
generations require ghosts-for-ghosts, so that new infinite towers of
ghosts-for-ghosts are involved.
%Finally, the counting of fields is not quite
%clear since the correspondence is not precise at the {\it n}'th
%generation level, because $\theta_1$ and $\tilthet_1$ are not the same.

{\it Acknowledgements}:
The author wishes to express his gratitude to C. M. Hull for encouragement,
helpful discussions and reading of the manuscript. I wish to thank to
M. B. Green and A. Mikovi\'c for useful conversations, and also to J. Hibbitt
for correcting errors present in an earlier draft. The research reported in
this letter has been supported by grant from CONACYT-MEXICO.
\refout
\bye
\end